\documentclass[eqsecnum,floats,aps,nofootinbib,showpacs]{revtex4}
\usepackage{latexsym}
\usepackage{amssymb}
\usepackage{ifpdf}
\ifpdf
\usepackage[pdftex]{hyperref}
\else
\usepackage{hyperref}
\fi

\renewcommand{\d}{\mathrm{d}}
\renewcommand{\l}{\left(}
\renewcommand{\r}{\right)}
\usepackage{amsmath, amsthm, amssymb}

\def\be{\begin{equation}}
\def\ee{\end{equation}}
\def\beq{\begin{equation*}}
\def\eeq{\end{equation*}}
\def\ba{\begin{aligned}}
\def\ea{\end{aligned}}

\def\ov{\overline}
\def\w{\wedge}

\begin{document}
\title{On the choice of coupling procedure for the Poincar{\'e} gauge theory of gravity}

\author {Marcin Ka\'zmierczak}
\email{marcin.kazmierczak@fuw.edu.pl}
\affiliation{Institute of Theoretical Physics, Uniwersytet Warszawski, Ho\.{z}a 69, 00-681 Warszawa, Poland} 

\begin{abstract}
The gauge approach to the theory of gravity has been widely discussed
as an alternative to standard general relativity. The Poincar{\'e}
group, as a symmetry group of all relativistic theories in the absence
of gravitation, constitutes the most natural candidate for a gauge
group. Although the Poincar{\'e} gauge theory of gravity has been
elaborated over the years and cast into a beautiful formal framework,
some fundamental problems have remained unsolved. One of them concerns the
inclusion of matter. The minimal coupling procedure, which is employed
in standard Yang--Mills theories, appears to be ambiguous in the case
of gravity. We propose a slight modification of this procedure, which removes the ambiguity. Our modification justifies
some earlier results concerning the consequences of the Poincar{\'e}
gauge theory of gravity. In particular, the predictions of Einstein--Cartan
theory with fermionic matter are rendered unique. 
We recall the earlier proposed solution based on modified volume--forms.
The advantage of our modification is that the predictions of the
theory are not radically changed. Basically, this modification simply
justifies the results that were obtained partly `by chance' in the
hitherto prevailing accounts on the Einstein--Cartan theory. The only
difference in the predictions, when compared to the standard treatment, concerns the Proca field in the
presence of gravity. The `torsion singularities' which occur there are
shifted towards other values of the field. 
\end{abstract}
\pacs{04.50.Kd, 04.40.-b, 11.30.Er, 11.15.-q}
\maketitle
\section{Introduction}\label{section1}
As noted by Einstein himself \cite{Eins}, the theory of gravity can be
formulated by considering the metric and the connection as
independent fields, which is referred to as first order formalism. It was originally observed by Weil \cite{Weil} that this formulation leads to different predictions,
when compared to the standard second order metric approach, after
fermions are included. At the same time, Weil pointed out that the second order
formalism would yield the results indistinguishable from those produced
by the first order one if fermionic Lagrangian was supplemented by an
additional term. However, this term would have to contain fourth order powers
of the Dirac field and the flat space fermionic Lagrangian thus
obtained would generate a nonlinear equation, instead of the standard Dirac one. It
is also important to stress that Weil concentrated on the
equations for the metric and the Dirac field. He would not be able to
state an
equivalence of the two formalisms if he attached any significance to the space--time
torsion.\par
Since the introduction by Yang and Mills of the non--Abelian gauge
theories \cite{YaMi}, attempts have been undertaken of describing all the known
interactions as emerging from the localization of some fundamental
symmetries of the laws of physics. It is now clear that all the
non--gravitational fundamental interactions can be successfully given
such an interpretation. The Yang--Mills theories constitute a formal basis
for the standard model of particle physics. However, gravitational
interaction has always been an odd one. Although the attempts to
describe gravity as a gauge theory were initiated by Utiyama \cite{Ut}
within a mere two years after the pioneering work of Yang and Mills,
the construction of this theory seems yet not to be satisfactorily
completed. All the relativistic theories in the absence of
gravity are invariant under the (global) action of the Lorenz group. They are
also trivially invariant under space--time translations. Therefore, it seems
natural to adopt either the Lorenz group or the full Poincar{\'e}
group as a gauge group. Utiyama's approach employed the Lorenz group,
but then it was necessary to introduce {\it ad hoc} a set of fields,
which were subsequently identified with the tetrad system on the
resulting Riemannian manifold. It was then observed by Kibble
\cite{Kib1} that this and other failings of Utiyama's theory can be remedied if the full Poincar{\'e} group is
promoted to the gauge group. This necessitates the first order
formalism for general relativity, with metric (but non--symmetric)
connection, as the set of Yang--Mills fields has to consist of ten
independent one--forms. At the same time, another aesthetic
deficiency of GR was removed. At the microscopic level, elementary
particle states can be classified by the irreducible unitary
representations of the (universal covering of the) Poincar{\'e}
group, which are labeled by mass and spin of the particle. It seems
rather artificial to assume that distribution of masses produces the curvature of
space--time, whereas spin does not influence the geometry
at all. In the Poincar{\'e} gauge theory, the macroscopically averaged
distribution of spin appears to be a source of the space--time
torsion. This fact was in agreement with the earlier ideas of Eli{\'e}
Cartan and hence the resulting modification of GR is usually referred to as
Einstein--Cartan theory. The curvature and the torsion of space--time are parts of the curvature of
the Poincar{\'e} connection of the underlying gauge
theory, which is explained in Section \ref{section2}.\par
If a field theory in Minkowski space is given, this theory being symmetric
under the global action of a representation of a Lie group, the
natural way to introduce the corresponding interaction within the
spirit of Yang--Mills theory is to apply the minimal coupling
procedure (MCP). Indeed, in the standard model of particle physics
this procedure is followed on the fundamental level, leading to predictions that agree with experimental results with great
accuracy. 
In GR, the principle of equivalence, which states that the
effects of gravitation can be locally `turned off' by a suitable
choice of a reference frame, necessitates minimal coupling. This
principle alone can be used to derive a majority of predictions of GR
(see \cite{WeinGr} and Appendix \ref{A2}), which are again excellently
confirmed by observations and experiments. However, trying to
apply MCP in order to pass from a field theory in flat space to
a Riemann--Cartan space--time
\footnote{A Riemann--Cartan space is a
  manifold with a metric tensor and a metric connection (in general
  non--symmetric).} results in difficulties. This is because
adding a divergence to the flat space Lagrangian density, which is a
symmetry transformation, leads to the non--equivalent
theory in curved space after MCP is applied. Although this problem was
observed already by Kibble, it has been largely ignored in the
subsequent investigations concerning EC theory. The paper \cite{HD}
provides an interesting example. The authors begin with a particular
flat space fermionic Lagrangian, which is quadratic in fields and
generates the Dirac equation. Then MCP is applied and standard
gravitational (first order) term is added to the Lagrangian. Owing to
the algebraic character of the equation that relates the density of spin
and the torsion, the latter can be expressed through matter fields
and the result can be inserted back into the Lagrangian. In this way
an effective action is obtained that depends on the metric and the
matter fields only. Its gravitational part is represented by the
standard second order action of GR. The matter part, however, differs from
the one that would be obtained via MCP in standard GR from the same initial flat space
fermionic Lagrangian. The difference is represented by an
additional term, which is the square of the Dirac axial
current, preceded by a very small coupling constant (of order $l^2$,
where $l$ is the Planck length). This term can be interpreted as
describing a gravitationally--induced point--interaction between
fermions. Further, one can find the
equation for the Dirac field, which is nonlinear, even in the
limit of the space--time metric being Minkowski's flat one. Although derivations presented in \cite{HD} were slightly
different from the line of reasoning presented above, the existence of such a weak contact interaction in EC
theory was the main result of the paper in question. But is it
justified to claim that the occurrence of this interaction
distinguishes the EC theory from GR? In fact, one could find the
Minkowskian limit of the matter part of the effective Lagrangian and
use the resulting modified fermionic Lagrangian as a starting point for
MCP in GR. One would than trivially obtain the same curved space
action as the effective one of EC theory. Hence, as far as torsion is
not given a fundamental significance, the two theories (EC and GR)
could be considered as indistinguishable. This is exactly what Weil
concluded already in \cite{Weil}. However, in GR we need to put the
point interaction into the theory by hand, whereas in EC it arises
naturally as a consequence of an interrelation between spin and
torsion. If we postulated that a simple (second order in field powers) Lagrangian generating
standard Dirac equation should be used as a starting point for MCP in
each case (one could discuss whether such a postulate is justified or
rather artificial), then it could seem to follow from the analyses of
\cite{Weil} and \cite{HD} that EC gravity indeed differs from GR by
the presence of the above discussed point fermion interaction.\par
Unfortunately, even such a statement, after all the stipulations that
we have made, would not be true. It is namely important which simple
Lagrangian generating Dirac equation is used. In flat space, we are
free to add a divergence of a vector field to the Lagrangian density,
which would make difference for the final theory in the
Riemann--Cartan space, as mentioned above. Exploiting this freedom,
one can produce, in addition to the axial--axial point--interaction, also
a vector--vector one, as well as a parity breaking axial--vector
one. The coupling constants of these interactions may reach
significant values. What is more, one can eliminate all of them on the effective level and thus render the predictions of EC theory with fermions indistinguishable
from the predictions of GR, in spite of all our limitations. These
facts were discussed in \cite{Kazm1}. Hence, it seems that MCP should
be somehow modified for the sake of connections with torsion, so that
it gives equivalent results for equivalent flat space
Lagrangians. This issue is of principal importance if EC theory is to
regain its predictive power. An attempt to establish such a modified
procedure was made by Saa \cite{Saa1}\cite{Saa2}. Unfortunately, Saa's
solution results in significant departures from standard GR, which
seem incompatible with observable data \cite{BFY}\cite{FY},
unless some additional assumptions of rather artificial nature are made, such as demanding
a priori part of the torsion tensor to vanish \cite{RMAS}. The main
purpose of this paper is to introduce an alternative modification of
MCP, which also eliminates the ambiguity. Unlike Saa's proposal, our
approach does not lead to radical changes in the predictions of the
theory. In the case of gravity with fermions, the procedure simply
justifies the earlier results of \cite{HD,HH,Ker,Rumpf,CI}. These results were
obtained partly `by chance', as the flat space Dirac Lagrangian was
randomly selected from the infinity of equally good possibilities. The
same applies to the paper by Perez and Rovelli \cite{PR} in which the
semi--classical physical effects of the Immirzi parameter of Loop
Quantum Gravity were discussed. As shown in \cite{Kazm2}, the
predictions of EC gravity with Holst term and fermions radically
change when we pass from one equivalent fermionic
Lagrangian to another. However, if we use the corrected coupling procedure
proposed here, instead of the standard MCP, the two--parameter family
of theories of \cite{Kazm2} shrinks into the unique theory, which is
exactly the one described in \cite{PR}.\par
After the Poincar{\'e} gauge theory was brought to life by
Kibble, the relation between the translational gauge
fields and the tetrad from the point of view of fibre bundle
formulation was clarified by Trautman \cite{T1} (see also \cite{SW}). There the
whole affine group was considered as a gauge group. The geometric
concept of a radius
mapping, which was used in  \cite{T1}, is
intimately connected with the Poincar{\'e} coordinates, whose physical
meaning was later elaborated in \cite{GN}. For an exhaustive review of possible
approaches to the formulation of gauge theory of gravity, see
\cite{HCMN}. In this paper we aim to formulate the theory in a possibly
simple (though precise) manner, rather than to acquire the highest degree of generality
and mathematical complexity. Such a simple formulation is presented in
Section \ref{section2}. In Section \ref{amb} we explain the origin and
consequences of the ambiguity of MCP in the presence of torsion, using
the Dirac field case as an instructive example. In Section \ref{ambr}
we present two possible modifications of MCP which
remove the ambiguity. First of them is the earlier Saa's procedure and
the second is our proposal. We comment on the aesthetical advantages
of a Poincar{\'e} gauge formulation of gravity combined with our coupling
procedure. In Section
\ref{cons} we describe the consequences of application of our
procedure to different field theories. An interesting issue is the
fate of torsion singularities in the theory of Proca field. Finally in
Section \ref{conc} we draw the conclusions.
\section{The Poincar{\'e} gauge theory}\label{section2}
\subsection{Yang--Mills theories}\label{YM}
Let us recall the classical formalism of a Yang--Mills gauge theory of
a Lie group $G$. Let
\be\label{S}
S[\phi]=\int\mathcal{L}\l{\phi,\partial_{\mu}\phi}\r
d^4x=\int\mathfrak{L}\l{\phi,d\phi}\r
\ee
represent the action of a field theory in Minkowski space $M$. Here
$\mathcal{L}$ is a Lagrangian density and $\mathfrak{L}$ a Lagrangian four--form. Assume
that $\mathcal{V}$ is a (finite dimensional) linear space in which fields $\phi$ take their values,
$\phi:M\rightarrow \mathcal{V}$, and $\pi$ is a representation of $Lie(G)$ on $\mathcal{V}$.
Let $\rho$ denote the corresponding representation of the
group\footnote{More precisely, in a generic case $\rho$ is a
  representation of the universal covering group of $G$, which may not
  be a representation of $G$ itself.},
$\rho\l{\exp (\mathfrak{g})}\r=\exp\l{\pi(\mathfrak{g})}\r$. Suppose
that the Lagrangian four--form, and hence the action, is invariant under its global action
\be\label{global}
\mathfrak{L}\l{\rho(g)\phi,d\l{\rho(g)\phi}\r}\r=\mathfrak{L}\l{\rho(g)\phi,\rho(g)d\phi}\r=
\mathfrak{L}\l{\phi,d\phi}\r .
\ee
Then one can introduce an interaction associated to the symmetry group
$G$ by allowing the group element $g$ to depend on space--time point
and demanding the theory to be invariant under the local action of
$G$. This can be most easily achieved by replacing the
differential by the covariant differential
\be\label{MCPYM}
d\phi\rightarrow D\phi=d\phi+\mathbb{A}\phi ,
\ee
where $\mathbb{A}$ is a $Lie(G)$--valued\footnote{More precisely,
  $\mathbb{A}$ and $\mathbb{F}$ take values in the representation
  $\pi$ of $Lie(G)$. The same concerns similar situations
  appearing later on.}  one--form field on $M$ which
transforms under the local action of $G$ as
\be\label{gauge}
\mathbb{A}\rightarrow\mathbb{A}'=\rho(g)\mathbb{A}\rho^{-1}(g)-d\rho(g)\rho^{-1}(g) .
\ee
Then $D'\phi'=\rho(g)D\phi$ and $\mathfrak{L}\l{\phi,D\phi}\r$ is
invariant under the local action of $G$, on account of (\ref{global}). If the
basis of $Lie(G)$ is chosen, the components of $\mathbb{A}$ represent
Yang--Mills fields of the resulting theory. The field strength can be
represented by a $Lie(G)$--valued two--form 
\beq
\mathbb{F}:=d\mathbb{A}+\mathbb{A}\mathbb{A} ,
\eeq
which transforms in the adjoint representation of $Lie(G)$ and obeys
Bianchi identity
\be\label{bian}
\mathbb{F}'=\rho(g)\mathbb{F}\rho^{-1}(g), \quad
d\mathbb{F}+[\mathbb{A},\mathbb{F}]=0 \ .
\ee
Think of $\mathbb{A}$ and $\mathbb{F}$ as matrices belonging to the
representation of $Lie(G)$ whose entries are one--forms on
$M$. When multiplying such matrices, one should multiply their entries
externally. The next step is the construction of a gauge--field part
of the action, which should be built of $\mathbb{A}$ and remain
invariant under gauge transformations (\ref{gauge}). 
\subsection{The Poincar{\'e} group as a gauge group}\label{PGGG}
The Poincar{\'e} group $\mathcal{P}$ consists of all isometries of Minkowski
space. They can be represented by pairs $(\Lambda,a)$ acting on $M$ as
$(\Lambda,a)x=\Lambda x+a$,
where $x\in M$, $a$ represents a column of four real numbers and
$\Lambda\in O(1,3)$ is a Lorenz matrix \footnote{We will not discuss
  parity and time reversal transformations in this article, hence one
  can think of $\Lambda$ as a proper ortochronus Lorenz matrix.}. One
easily finds out that the composition law is
$(\Lambda_1,a_1)(\Lambda_2,a_2)=(\Lambda_1\Lambda_2,\Lambda_1a_2+a_1)$.
Let 
\beq
\ba
&\rho(\Lambda,a):=\rho(a)\rho(\Lambda),\\
&\rho(a):=\exp\l{a_{a}P^a}\r,\quad 
\rho\l{\Lambda(\varepsilon)}\r:=\exp\l{\frac{1}{2}\varepsilon_{ab}J^{ab}}\r
\ea
\eeq
be the representation of $\mathcal{P}$. Here $P^a, J^{ab}$ are the
generators of translations and Lorenz rotations and belong to the
representation $\pi$ of $Lie(\mathcal{P})$ (see previous
subsection). The coefficients $\varepsilon_{ab}=-\varepsilon_{ba}$ are
parameters of the Lorenz transformation 
$\Lambda(\varepsilon)=\exp (\varepsilon)$. Here $\varepsilon\in so(1,3)$
is the matrix with entries
${\varepsilon^a}_b:=\eta^{ac}\varepsilon_{cb}$, where
$(\eta^{ab})=(\eta_{ab})=diag(1,-1,-1,-1)$ is Minkowski matrix. Using
the composition law and employing infinitesimal transformations one
can derive transformation properties of the generators and the commutation relations for the
Poincar{\'e} algebra (see \cite{Wein} for these derivations)
\be\label{transcom}
\ba
&\rho(\Lambda,a)P^a\rho^{-1}(\Lambda,a)={\Lambda_c}^aP^c,\\ 
&\rho(\Lambda,a)J^{ab}\rho^{-1}(\Lambda,a)={\Lambda_c}^a{\Lambda_d}^b\l{J^{cd}+a^cP^d-a^dP^c}\r , \\
&[P^a,J^{cd}]=\eta^{ac}P^d-\eta^{ad}P^c, \\ 
&[P^a,P^b]=0, \\
&[J^{ab},J^{cd}]=\eta^{ad}J^{bc}+\eta^{bc}J^{ad}-\eta^{bd}J^{ac}-\eta^{ac}J^{bd} .
\ea
\ee
If gravity is not present, special theory of relativity forces all the
field theories to be invariant under the global action of the
Poincar{\'e} group. It would be a tempting idea to derive gravitational
interaction by demanding this symmetry to hold locally. Then, gravity
would emerge from special relativity in much the same way as
electromagnetism emerges from the invariance of a matter Lagrangian
under the change of phase. In order to construct the covariant
differential, one needs to introduce the $Lie(\mathcal{P})$-valued one--form
\be\label{pA}
\mathbb{A}=\frac{1}{2}\omega_{ab}J^{ab}+\Gamma_aP^a ,
\ee
where $\omega_{ab}=-\omega_{ba}$ and $\Gamma_a$ are
one--forms\footnote{Here $\mathcal{M}$ is the space--time manifold,
  which will no longer be the Minkowski space $M$ in the presence of gravity.} on
$\mathcal{M}$. Under gauge transformations (\ref{gauge}), these one--forms
transform as
\be\label{gftrans}
\omega'=\Lambda\omega\Lambda^{-1}-d\Lambda\Lambda^{-1}, \quad
\Gamma'=\Lambda\Gamma-\omega'a-da ,
\ee
which can be easily verified for infinitesimal transformations (use
(\ref{transcom})). Here $\omega$ is a matrix with entries
${\omega^a}_b$ and $\Gamma$ a column matrix with entries
$\Gamma^a$. It is now time to recall that we aim to formulate the
theory of gravitational interaction, which ought to be bound up with
the geometry of space--time, according to Einstein's idea. The first
equation of (\ref{gftrans}) is simply the transformation rule for
connection one--forms under the change of an orthonormal
frame of vector fields. Indeed, orthonormal frames on Lorenzian manifold are connected
by local Lorenz transformations. Note also that the antisymmetry
$\omega_{ab}=-\omega_{ba}$ means metricity of the resulting
space--time connection. Unfortunately, we do not have a metric given
{\it apriori} on space--time, after we have disposed of the flat
Minkowski's one. Hence, we need to provide somehow the space--time
with metric structure, preferably via the introduction of a cotetrad
field $e$ (it has long been known that tetrads are necessary to include
fermions in the theory of gravity \cite{Weil}). The transformation
formula for $e$, compatible with the one for $\omega$ (\ref{gftrans}),
would be $e'=\Lambda e$ (think of $e$ as a column of one--forms
$e^a=e^a_{\mu}dx^{\mu}$). Therefore one cannot just adopt the
translational gauge field $\Gamma$ as representing cotetrad. The
solution is to introduce a vector--valued zero--form $y$ (a column of
functions on $\mathcal{M}$), which transformas under the local
Poincar{\'e} transformation $(\Lambda,a)$ according to
\be\label{ytrans}
y'=\Lambda y+a ,
\ee
 and then introduce the cotetrad
\be\label{e}
e:=\Gamma+Dy , \qquad Dy=dy+\omega y .
\ee
$D$ will always denote the Lorenz covariant derivative (see
Appendix \ref{A1}). Then from (\ref{gftrans}) and (\ref{ytrans}) it follows that
$e'=\Gamma'+dy'+\omega'y'=\Lambda e$, as desired. Although the new
field $y$ can seem to have been introduced {\it ad hoc}, it can be given a natural geometric interpretation in the language of fibre
bundles \cite{T1}, as well as the physical meaning
\cite{GN}\cite{T3h}\cite{Lh}. What is more, if the Lagrangian
four--form depends on $y$ and $\Gamma$ only via the cotetrad $e$, one
is free to acknowledge $e$ as a fundamental field and forget about its
origin. Indeed, the variation of such a Lagrangian would be
\be\label{ew}
\delta{\mathfrak{L}}=
\delta e^a\wedge\frac{\delta\mathfrak{L}}{\delta e^a}+
\delta\omega^{ab}\wedge\frac{\delta\mathfrak{L}}{\delta\omega^{ab}}+
\delta\phi\wedge\frac{\delta\mathfrak{L}}{\delta\phi}  ,
\ee
$\phi$ representing matter fields. Since $\delta
e^a=\delta\Gamma^a+D\delta y^a+\delta{\omega^a}_by^b$, we finally get
\be\label{gyw}
\ba
&\delta{\mathfrak{L}}=
\delta \Gamma^a\wedge\frac{\delta\mathfrak{L}}{\delta e^a}-
\delta y^a D\l{\frac{\delta\mathfrak{L}}{\delta e^a}}\r\\
+&\delta\omega^{ab}\wedge\l{y_b\frac{\delta\mathfrak{L}}{\delta e^a}+\frac{\delta\mathfrak{L}}{\delta\omega^{ab}}}\r+
\delta\phi\wedge\frac{\delta\mathfrak{L}}{\delta\phi}+d\l{\delta y^a\frac{\delta\mathfrak{L}}{\delta e^a}}\r .
\ea
\ee
Comparing (\ref{gyw}) with (\ref{ew}) one can see that promoting $e$
to the fundamental field, instead of $\Gamma$ and $y$, do not
influence the resulting system of field equations. Having the cotetrad introduced, we
can define the {\it torsion two--form} $Q^a:=De^a=\frac{1}{2}{T^a}_{bc}e^b\wedge e^c$. Then using (\ref{pA})
and relations (\ref{transcom}) one finds that
\beq
\mathbb{F}=\frac{1}{2}\Omega_{ab}J^{ab}+\l{Q_a-{\Omega_a}^by_b}\r P^a ,
\eeq
where ${\Omega^a}_b:=d{\omega^a}_b+{\omega^a}_c\wedge{\omega^c}_b=\frac{1}{2}{R^a}_{bcd}e^c\wedge e^d$ is the {\it
  curvature two--form}. One can check that $\mathbb{F}$ obeys
(\ref{bian}) (the Bianchi identity appears to be equivalent to
$d\Omega+[\omega,\Omega]=0, \ DQ=\Omega e$).\par
By adopting
$\mathfrak{L}=\mathfrak{L}_G+\mathfrak{L}_m$ as a Lagrangian, where 
$\mathfrak{L}_G=-\frac{1}{4k}\epsilon_{abcd}e^a\wedge e^b\wedge
\Omega^{cd}$ represents gravitational part and $\mathfrak{L}_m$ the matter
part (here $k=8\pi G$, where $G$ is the gravitational constant), one
recovers the field equations of the Einstein--Cartan theory of gravity
\be\label{FEq}
\ba
&\frac{\delta\mathfrak{L}_G}{\delta e^a}+\dfrac{\delta\mathfrak{L}_m}{\delta e^a}=0  \ &\Leftrightarrow& \ 
&{G^a}_b:={R^a}_b-\frac{1}{2}R\delta^a_b=k\,{ t_b}^a \\
&\frac{\delta\mathfrak{L}_G}{\delta\omega^{ab}}+\dfrac{\delta\mathfrak{L}_m}{\delta\omega^{ab}}=0 \ &\Leftrightarrow& \
&T^{cab}-T^a\eta^{bc}+T^b\eta^{ac}=k S^{abc}  \\
&\frac{\delta\mathfrak{L}_m}{\delta\phi}=0
\ea
\ee
where ${R^a}_b:=\eta^{ac}{R^d}_{cdb}$, $R:={R^a}_a$, $T^a:={T^{ba}}_
b$ and the dynamical definitions of energy--momentum and spin density tensors on
Riemann--Cartan space are
\be\label{ts}
t_{ab}e^b:=-\star\dfrac{\delta\mathfrak{L}_m}{\delta e^a}, \quad
 S^{abc}e_c:=2\star\dfrac{\delta\mathfrak{L}_m}{\delta\omega_{ab}}.
\ee
Here $\star$ is the Hodge star of the cotetrad--induced metric
$g=\eta_{ab}e^a\otimes e^b$ (see Appendix \ref{A1}). We have not yet explained how to construct the matter Lagrangian
$\mathfrak{L}_m$ from its flat space counterpart. This issue appears
to be problematic and will be
considered in the following section.
\section{Coupling gravity to matter fields -- standard approach}\label{amb}
Let (\ref{S}) denote the action functional of a classical field theory
in Minkowski space $M$.
It is well known that the transformation
\be\label{Lch}
\mathcal{L}\rightarrow\mathcal{L}'=\mathcal{L}+\partial_{\mu}V^{\mu}
\ee
of the Lagrangian density changes $\mathfrak{L}$ by a differential
\be\label{differential}
\partial_{\mu}V^{\mu}\, d^4x=\pounds_V\,d^4x=-\star d x_{\mu}
\wedge d V^{\mu}=d (V \lrcorner \, d^4 x)  ,
\ee
where $\pounds$ denotes the Lie derivative, $\lrcorner$ the internal
product, $\star$ is the Hodge star of the flat Minkowski metric and
$x^{\mu}$ are inertial coordinates on $M$. Not only does this
transformation not change the field equations
generated by $S$, but it also leaves the integrated
energy and momenta obtained via Noether procedure invariant
\cite{Kazm1}. Despite the more subtle behavior of the spin density
tensor \cite{Kazm1}, the transformation still seems to be a true
symmetry of the theory. We wish now to introduce a new interaction. A
reasonable consistency condition would be to require the
resulting theory not to depend on whether we have added a divergence
to the initial Lagrangian density or not. For the sake of simplicity, let us consider an example of the Dirac field. The most
frequently used Lagrangian four--form is
\be\label{LF0}
\ba
&\mathfrak{L}_{F0}=
-i\l{\star dx_{\mu}}\r \wedge \ov{\psi}\gamma^{\mu}
  d\psi-m\ov{\psi}\psi\, d^4x \\
&=\ov{\psi}\l{i\gamma^{\mu}\partial_{\mu}-m}\r\psi\, d^4x . 
\ea
\ee
Here $\gamma^{\mu}$ are the Dirac matrixes obeying 
$\gamma^{\mu}\gamma^{\nu}+\gamma^{\nu}\gamma^{\mu}=2\eta^{\mu\nu}$ and
$\ov{\psi}:={\psi}^{\dagger}\gamma^0$, where ${\psi}^{\dagger}$ is
a Hermitian conjugation of a column matrix (think of $\psi$ as a
column of four complex--valued functions on space--time).
In the spirit of the conventional Yang--Mills theory, we should merely
perform the replacement (\ref{MCPYM}) to `turn on' the interaction. Explicitly, 
\be\label{MCPpsi}
d\psi\rightarrow d\psi+\mathbb{A}\psi,\quad
d\ov{\psi}\rightarrow d\ov{\psi}+\ov{\psi}\gamma^0\mathbb{A}^{\dagger}\gamma^{0}.
\ee
Consider addition of
a divergence of a vector field of the form
\be\label{V}
V^{\mu}=\ov{\psi}B^{\mu}\psi, \quad B^{\mu}=a\gamma^{\mu}+b\gamma^{\mu}\gamma^5
\ee
to the initial Lagrangian density. Here $a$ and $b$ are arbitrary
complex numbers.
 This form was chosen so
that the new Lagrangian was equally `reasonable' as the original one
(quadratic in fields, invariant under global proper Poincar{\'e}
transformations).Under the replacement (\ref{MCPpsi}), the differentials $dV^{\mu}$ transform as
\be\label{dV}
d\ov{\psi}B^{\mu}\psi+\ov{\psi}B^{\mu}d\psi\rightarrow 
dV^{\mu}+\ov{\psi}\l{\gamma^0\mathbb{A}^{\dagger}\gamma^0B^{\mu}+B^{\mu}\mathbb{A}}\r\psi.
\ee 
In the case of non--gravitational interactions, the gauge groups are
unitary (hence $\mathbb{A}$ is anty--hermitian) and their
representations do not act on the spinor indedices of $\psi$, which means
that $\mathbb{A}$ commutes with $\gamma^0$ and $B^{\mu}$. Next, using $\gamma^0\gamma^0=1$, we can conclude that $dV^{\mu}$ remains
unchanged and (\ref{differential}) is still a differential. Hence, the two
equivalent non--interacting theories give rise to the equivalent
theories with interaction.\par
Let us now turn to gravity. The Lagrangian four--form (\ref{LF0}) is
invariant under the global action of the Poincar{\'e} group
\beq
\ba
&x\rightarrow x'=\Lambda x+a, \quad \psi\rightarrow {\psi}'=S(\Lambda)\psi, \\
&S(\Lambda(\varepsilon)):=\exp\l{-\frac{i}{4}\varepsilon_{\mu\nu}\Sigma^{\mu\nu}}\r,
 \ \Sigma^{\mu\nu}:=\frac{i}{2}[\gamma^{\mu},\gamma^{\nu}] 
\ea
\eeq
(use the identity
$\gamma^0S^{\dagger}(\Lambda)\gamma^0=S^{-1}(\Lambda)$ to check this
invariance). In order to make this symmetry local, it is not sufficient to perform
the substitution (\ref{MCPpsi}). Now we pass from
Minkowski space $M$ to the Riemann--Cartan manifold
$\mathcal{M}(\omega,e)$ -- the manifold with the metric structure
(described by $e$) and the metric--compatible connection. We should therefore
replace the basis of one--forms $dx^{\mu}$ of $M$ by the cotetrad
basis $e^a$ of $\mathcal{M}$ and use the Hodge star operator $\star$ adapted to
$\mathcal{M}$. The resulting Lagrangian four--form is
\be\label{LF0gr}
\ba
&\tilde{\mathfrak{L}}_{F0}=
-i\l{\star e_a}\r \wedge \ov{\psi}\gamma^aD\psi-m\ov{\psi}\psi\,\epsilon , \\
&D\psi=d\psi-\frac{i}{4}\omega_{ab}\Sigma^{ab}\psi 
\ea
\ee
(the matrixes $\gamma^a$, $a=0,\dots,3$ are just the same as
$\gamma^{\mu}$, $\mu=0,\dots,3$). Here $\epsilon=e^0\wedge e^1\wedge
e^2\wedge e^3$ is the canonical volume element on $\mathcal{M}$.\par
In more general terms, for a flat space field theory of a field $\phi$, which is to be adapted to
the Riemann--Cartan manifold $\mathcal{M}$ with the metric $g$ and the
metric--compatible connection $\nabla$, the procedure thus described amounts to the
passage
\be\label{MCP}
\mathcal{L}(\phi,\partial_{\mu}\phi,\dots )\,\d^4 x 
\longrightarrow  \mathcal{L}(\phi,\nabla_{\mu}\phi,\dots
)\,\epsilon ,
\ee
where $\epsilon=\sqrt{|\det g|} \, d^4 x$ is the canonical volume
four--form, $\det g$ being the determinant of the matrix of components
$g_{\mu\nu}=g(\partial_{\mu},\partial_{\nu})$ of the metric tensor. The dots
correspond to the possibility of $\mathcal{L}$ to depend on
higher derivatives of fields. Here $\nabla_{\mu}\phi$ denotes the
appropriately defined covariant derivative of $\phi$ with respect to
the connection $\nabla$ of $\mathcal{M}$ (the details depend on the
particular field $\phi$). This procedure will be
referred to as the minimal coupling procedure (MCP) for the
gravitational interaction.\par
From the form of the covariant derivative $D\psi$ in (\ref{LF0gr}) one can read out the
generators of the relevant representation $\rho$ of the Poincar{\'e} group
and find the representation itself
\beq
P^a=0,\quad J^{ab}=-\frac{i}{2}\Sigma^{ab}, \quad
\rho(\Lambda,a)=S(\Lambda) .
\eeq
One can find out that they satisfy the relations (\ref{transcom}) and
that $D'\psi'=S(\Lambda)D\psi$, on account of the first equation of
(\ref{gftrans}). Hence, the Lagrangian four--form (\ref{LF0gr}) is
invariant under the local action of the gauge group, as desired. In
the most conventional approach to the Poincar{\'e} gauge theory of
gravity with fermions, which we follow in this article, the
translational part of the Poincar{\'e} group is realized
trivially. The possibility and consequences of a nontrivial
realization of translations 
were addressed in \cite{GN}\cite{Lec1}\cite{TT1}\cite{TT2}.\par
Let us now consider the effect of the transformation (\ref{Lch}),
performed on the initial Lagrangian, on the final Lagrangian
four--form on Riemann--Cartan manifold. We shall consider the vector
field of the form (\ref{V}).
It is straightforward to check that the following Leibniz rule applies
\be\label{Leib}
\l{D\ov{\psi}}\r
B^a\psi+\ov{\psi}B^aD\psi=d\l{\ov{\psi}B^a\psi}\r+{\omega^a}_b \l{\ov{\psi}B^b\psi}\r
\ee
(use $[\gamma^a,\Sigma^{bc}]=4i\eta^{a[b}\gamma^{c]}$). One can next decompose the
connection $\omega$ into the Levi--Civita part and the part determined
by torsion (see \cite{Kazm1} for this kind of calculations) and
finally conclude, that the change in the resulting Lagrangian four--form on
$\mathcal{M}$ will be
\be\label{TVtetr}
d\l{V\lrcorner\epsilon}\r-T_aV^a\epsilon ,
\ee
where $T^a$ is the torsion trace vector introduced in
(\ref{FEq}). Within the framework of classical general relativity,
where the torsion of the connection is assumed to vanish, the
result would be again a differential. In Einstein--Cartan theory the
torsion is determined by the spin of matter via the second equation of
(\ref{FEq}) and does not vanish in general. Hence, the equivalent
theories of the Dirac field in flat space can lead to the
non--equivalent theories with gravitation.\par
Let us ignore this problem for a moment and look at the field equations
generated by (\ref{LF0gr}) resulting from variation with respect to
$\psi$ and $\ov{\psi}$. The variational derivatives are
\beq
\ba
&\frac{\delta\tilde{\mathfrak{L}}_{F0}}{\delta\ov{\psi}}=\l{i\gamma^aD_a\psi-m\psi}\r\epsilon,\\
&\frac{\delta\tilde{\mathfrak{L}}_{F0}}{\delta{\psi}}=
\l{-i\gamma^a\ov{D_a\psi}-m\ov{\psi}-iT_a\ov{\psi}\gamma^a}\r\epsilon,
\ea
\eeq
where $D_a\psi$ denotes the $a$-th component of a one--form $D\psi$ in
the cotetrad basis: $D\psi=\l{D_a\psi}\r e^a$. To derive this results,
it is useful to know the identity $D\l{\star
  e_a}\r=T_{a}\epsilon$. Contrary to what would be expected from the
Dirac field, calculating the variational derivative with respect to
$\ov{\psi}$ and equating it to zero yields the equation that is not equivalent to the one
obtained by varying with respect to $\psi$. For $\mathfrak{L}_{F0}$, the
equivalence of the corresponding equations follows from the fact that
$\mathfrak{L}_{F0}$ differs by divergence from the real Lagrangian four--form
\be\label{LFR}
\mathfrak{L}_{FR}=
-\dfrac{i}{2}\l{\star dx_{\mu}}\r \wedge \l{ \ov{\psi}\gamma^{\mu}
  d\psi-\ov{d\psi}\gamma^{\mu}\psi}\r-m\ov{\psi}\psi d^4x .
\ee
But this is no longer the case for $\tilde{\mathfrak{L}}_{F0}$. The
commonly accepted solution to this problem is to adopt (\ref{LFR}) as
an appropriate flat space Lagrangian. Then the application of MCP yields
\beq
\tilde{\mathfrak{L}}_{FR}=
-\dfrac{i}{2}\l{\star e_a}\r \wedge \l{ \ov{\psi}\gamma^a
  D\psi-\ov{D\psi}\gamma^a\psi}\r-m\ov{\psi}\psi\,\epsilon  
\eeq
and the field equations obtained by varying with respect to $\ov{\psi}$
and $\psi$ are
\beq
\ba
&i\gamma^aD_a\psi-m\psi+\frac{i}{2}\gamma^aT_a\psi=0,\\
&-i\gamma^a\ov{D_a\psi}-m\ov{\psi}-\frac{i}{2}T_a\ov{\psi}\gamma^a=0 .
\ea
\eeq
Hence, they are equivalent. This choice of Lagrangian served as the
basis for physical investigations in numerous papers. But the reality
requirement does not fix the theory uniquely. We can next add to
$\mathcal{L}_{FR}$ the
divergence of a vector field of the form (\ref{V}), where now the parameters
$a$, $b$ are required to be real, since we do not want to destroy
the reality of the Lagrangian. If we required the Lagrangian to be parity
invariant, we would have to set $b=0$. But there are no arguments in
favor of choosing a particular value of $a$, except for some
speculations concerning the resulting form of the spin density tensor
\cite{Kazm1}. The nonzero values of the parameters $a$ and $b$ can
lead to the meaningful physical effects \cite{Kazm1}\cite{Kazm2}.\par
Hence, the standard MCP for the Poincar{\'e} gauge theory
of gravity appears to involve an ambiguity. In the next section
we will present possible solutions to the problem.
\section{How to remove the ambiguity?}\label{ambr}
\subsection{Preliminary remarks}\label{pre}
Let $\tilde{\mathcal{L}}$ and $\tilde{\mathcal{L}'}$ denote the
results of application of MCP (\ref{MCP}) to the flat space Lagrangian densities $\mathcal{L}$ and
$\mathcal{L}'$, connected by the transformation (\ref{Lch}). We could
expect their difference to be
\be\label{dif1}
\tilde{\mathcal{L}'}-\tilde{\mathcal{L}}=\nabla_{\mu}V^{\mu},
\ee
where $\nabla_{\mu}V^{\nu}$ is the standard abbreviation for the $\nu$--th
component of the covariant derivative of $V$ in the direction of a basis vector
field $\partial_{\mu}$, 
$
\nabla_{\mu}V^{\nu}:=(\nabla_{\partial_{\mu}}V)^{\nu}=\partial_{\mu}V^{\nu}+{\Gamma^{\nu}}_{\rho\mu}V^{\rho}
$,
where the connection coefficients in the holonomic basis $\partial_{\mu}$ are defined by
$\nabla_{\partial_{\mu}}\partial_{\nu}={\Gamma^{\rho}}_{\nu\mu}\partial_{\rho}$.
Similarly, the connection coefficients in the tetrad basis are provided by
$\nabla_{\tilde{e}_a}\tilde{e}_b={\Gamma^c}_{ba}\tilde{e}_c$ and their
relation to the connection one--forms is ${\omega^a}_b={\Gamma^a}_{bc}e^c$.
The difference in the corresponding Lagrangian
four--forms can be expressed in terms of the cotetrad and the
connection one--forms as
\be\label{dif2}
\tilde{\mathfrak{L}'}-\tilde{\mathfrak{L}}=-\l{\star{e}_a}\r\wedge DV^a,
\ee
where $DV^a=dV^a+{\omega^a}_b V^b$ is the usual covariant
derivative of a differential zero--form of vectorial type (see
Appendix \ref{A1}).\par
Note that these equivalent statements (\ref{dif1}) and (\ref{dif2})
\footnote{Use $DV^a=(\nabla_bV^a) e^b$ and $(\star e_a)\wedge
e^b=-\delta^b_a\epsilon$ to show this equivalence.}
are not so obvious. One should think of $V$ as composed of the
fields of the theory, an example of such reasonable composition for
the Dirac theory being provided by (\ref{V}). In order to see what the
divergence of $V$ would become after the application of MCP, we have
to rewrite it in such a way that the differential
operators act directly on fields. Then the derivatives (or
differentials) of fields
should be replaced by the covariant ones. In the case of the Dirac
field, (\ref{dif2}) will be true for the vector field of the form (\ref{V}). This is because
the Leibniz rule (\ref{Leib}) holds in this case. In general, this rule would not
apply to $V^a=\ov{\psi}A^a\psi$, with the matrixes $A^a$ being
different from $B^a$ of (\ref{V}). But such $V^a$ would not represent
a genuine vector field, as its
components would transform improperly under Lorenz transformations. In the following,
we will only consider genuine vector fields, with correct
transformation properties, to which the results (\ref{dif1}) and
(\ref{dif2}) apply.\par
One can prove that 
\be\label{TV}
\nabla_{\mu}V^{\mu}={\stackrel{\circ}{\nabla}}_{\mu}V^{\mu}-T_{\mu}V^{\mu} ,
\ee
where $\stackrel{\circ}{\nabla}$ is the torsion--free Levi--Civita connection and
$T_{\mu}={T^{\nu}}_{\mu\nu}$ the torsion trace vector (compare with
(\ref{TVtetr})).\footnote{The torsion tensor can be defined in terms
  of the connection $\nabla$ by $T(X,Y)=\nabla_XY-\nabla_YX-[X,Y]$, where $X$, $Y$
  are vector fields and $[,]$ is the Lie bracket. In the holonomic basis,
the components are expressed through the connection coefficients by
${T^{\rho}}_{\mu\nu}=dx^{\rho}\l{T(\partial_{\mu},\partial_{\nu})}\r=-{\Gamma^{\rho}}_{\mu\nu}+{\Gamma^{\rho}}_{\nu\mu}$.
The components in the tetrad basis ${T^a}_{bc}=e^{a}\l{T(\tilde{e}_b,\tilde{e}_c)}\r$
coincide with those of the torsion two--form $Q^a$ introduced in
Section \ref{PGGG}.} 
\subsection{Modified volume form approach --  Saa's proposal}
After multiplied by the canonical volume element, the first term
in (\ref{TV}) becomes a Lie derivative of a four--form (and hence a differential) ${\stackrel{\circ}{\nabla
  }}_{\mu}V^{\mu}\epsilon=\pounds_V\epsilon=\d (V \lrcorner
\epsilon)$. This is not the case for the second term. The
basic idea of \cite{RMAS} is that we could make the whole expression
to be a Lie derivative, if we used another volume element $\tau=f\epsilon$,
instead of the canonical one. Here, $f$ is a nowhere vanishing
function, which should be suitably adapted to the
connection (all volume forms on the manifold differ by a nowhere vanishing function). Let us impose the requirement
\be\label{SaaCond}
\nabla_{\mu}V^{\mu}\tau=\pounds_V\tau .
\ee
On account of  (\ref{TV}), the LHS can be rewritten as 
$f{\stackrel{\circ}{\nabla}}_{\mu}V^{\mu}\epsilon-fT_{\mu}V^{\mu}\epsilon$,
whereas the RHS can be rewritten as
$\pounds_V(f\epsilon)=V(f)\epsilon+f\pounds_V\epsilon$. Since 
${\stackrel{\circ}{\nabla}}_{\mu}V^{\mu}\epsilon=\pounds_V\epsilon$
and $V(f)=V^{\mu}\partial_{\mu}f$, we can see that (\ref{SaaCond})
will be satisfied if and only if $T_{\mu}=-\partial_{\mu}\ln f$. The
solution to the problem of nonuniqueness of MCP procedure could
therefore be to postulate that the torsion trace should be derivable
from the potential $-\ln f$ and to promote this potential to the fundamental
field of the theory. At the same time, the coupling procedure
(\ref{MCP}) should be slightly modified - instead of the canonical
volume form, the connection compatible volume form $\tau=f\epsilon$
ought to be applied. This removes the ambiguity. However, the price is
considerably high, as the resulting field equations differ
significantly from the usual equations of general relativity, or
Einstein--Cartan gravity. Although a lot of interesting
effects can be observed on the ground of this theory, such as propagating torsion or coupling gauge fields to
torsion without breaking the gauge symmetry \cite{Saa1}\cite{Saa2},
some implications of the model seem not to be compatible with
observational data \cite{BFY}\cite{FY}. The solution could be to
postulate the torsion trace {\it a priori} to vanish \cite{RMAS}, but
such an artificial assumption significantly decreases the elegance of
the theory.
\subsection{Modified connection approach}\label{MCA}
Let us recall that the main purpose of the introduction of a
covariant derivative (\ref{MCPYM}) with $\mathbb{A}$ transforming
according to (\ref{gauge}) was to localize the symmetry. If we used the
modified covariant derivative 
\be\label{modcd}
\stackrel{\mathbb{B}}{D}\phi=D\phi+\mathbb{B}\phi,
\ee
with $\mathbb{B}$ transforming according to
\be\label{Btrans}
\mathbb{B}'=\rho(g)\mathbb{B}\rho^{-1}(g),
\ee
its transformation properties would remain correct, 
$(\stackrel{\mathbb{B}}{D}\phi)'=\rho(g)\stackrel{\mathbb{B}}{D}\phi$,
and the local symmetry would be preserved. In general, $\mathbb{B}$ is a
$Lin(\mathcal{V})$--valued one--form on
space--time, where $Lin(\mathcal{V})$ is a set of linear maps of
$\mathcal{V}$ into itself, $\mathcal{V}$ being a linear space in which $\phi$
takes its values. Such an additional one--form $\mathbb{B}$ appears
when a new gauge symmetry is introduced, in addition to the one
corresponding to the representation $\rho$. Then $\mathbb{B}$
transforms inhomogenously, according to (\ref{gauge}), under the
action of its own gauge group. Under the action of $\rho$,
$\mathbb{B}$ has to transform according to (\ref{Btrans}), since we do
not want to destroy the already existing gauge symmetry. 

Our aim is not to introduce a new gauge symmetry, but to
exploit the freedom of addition of $\mathbb{B}$ in the Poincar{\'e}
gauge theory, in order to modify
coupling procedure itself. The idea is to construct $\mathbb{B}$ from
$\mathbb{A}$ in such a way that the transformation law (\ref{Btrans})
follows from the one for $\mathbb{A}$ and the resulting modified
coupling procedure, emloying (\ref{modcd}), is free of the
ambiguity. We shall now consider the Dirac field case and we will see that under
some reasonable restrictions the solution to the above formulated
problem is unique. The most general $gl(4,\mathbb{C})$--valued
one--form can be expanded in the basis of matrixes ${\bf 1}$,
$\gamma^5$, $\gamma^a$, $\gamma^5\gamma^a$, $\Sigma^{ab}$,
\beq
\mathbb{B}=\chi{\bf 1}+\kappa\gamma^5+\tau_a\gamma^a+\rho_a\gamma^5\gamma^a+\frac{1}{2}\alpha_{ab}\Sigma^{ab},
\eeq 
where $\chi$, $\kappa$, $\tau_a$, $\rho_a$, $\alpha_{ab}=-\alpha_{ba}$ are complex
valued one--forms on space--time. Since we do not want to change the
original Poincar{\'e} gauge fields $\omega_{ab}$, which are contained
in $\mathbb{A}=-\frac{i}{4}\omega_{ab}\Sigma^{ab}$, we will require
\be\label{znikanie}
Im(\alpha_{ab})=0.
\ee
As noted in Section \ref{pre}, it is important to demand that the
covariant derivative obey the Leibniz rule while acting on vector
fields composed of spinor fields. We shall require this rule to hold
for vector and axial Dirac currents $J_{(V)}^a=\ov{\psi}\gamma^a\psi$,
$J_{(A)}^a=\ov{\psi}\gamma^a\gamma^5\psi$.
Explicitely, we will require that
\beq
\ba
&(\stackrel{\mathbb{B}}{D}\ov{\psi})\gamma^a\psi+\ov{\psi}\gamma^a\stackrel{\mathbb{B}}{D}\psi=
dJ_{(V)}^a+\stackrel{\mathbb{B}}{\omega}{^a}_bJ_{(V)}^b,\\
&(\stackrel{\mathbb{B}}{D}\ov{\psi})\gamma^a\gamma^5\psi+\ov{\psi}\gamma^a\gamma^5\stackrel{\mathbb{B}}{D}\psi=
dJ_{(A)}^a+\stackrel{\mathbb{B}}{\omega}{^a}_bJ_{(A)}^b,
\ea
\eeq
where
$\stackrel{\mathbb{B}}{D}\ov{\psi}=(\stackrel{\mathbb{B}}{D}\psi)^{\dag}\gamma^0$
and $\stackrel{\mathbb{B}}{\omega}{^a}_b$ represents a modified
connection on the Riemann--Cartan manifold, which will be
determined. These equations can be rewritten as
\be\label{leibB}
\ba
&\gamma^0\mathbb{B}^{\dag}\gamma^0\gamma^a+\gamma^a\mathbb{B}={\beta^a}_b\gamma^b,\\
&\gamma^0\mathbb{B}^{\dag}\gamma^0\gamma^a\gamma^5+\gamma^a\gamma^5\mathbb{B}={\beta^a}_b\gamma^b\gamma^5,
\ea
\ee 
where $\beta=\stackrel{\mathbb{B}}{\omega}-\omega$ is the difference
between the two connections on $\mathcal{M}$. Multiplying the first of
these equations by $\gamma^5$ and comparing with the second we can
conclude that $[\mathbb{B},\gamma^5]=0$ and hence
\beq
\tau_a=\rho_a=0.
\eeq
After this requirement is imposed, the second equation of
(\ref{leibB}) follows from the first one, so we will further deal with
the first one only. For $a=0$ it yields
\beq
\l{\chi+\chi^*}\r\gamma^0-\l{\kappa+\kappa^*}\r\gamma^5\gamma^0+i\l{\alpha_{0j}-\alpha_{0j}^*}\r\gamma^j+
\frac{1}{2}\l{\alpha_{ij}+\alpha_{ij}^*}\r\Sigma^{ij}\gamma^0={\beta^0}_b\gamma^b.
\eeq
Knowing that $\Sigma^{ij}\gamma^0=\varepsilon^{ijk}\gamma^5\gamma^k$,
where $\varepsilon^{123}=1$, we can conclude that 
\beq
{\beta^0}_0=2Re\l{\chi}\r,\quad Re(\kappa)=0,\quad
{\beta^0}_j=-2Im(\alpha_{0j}), \quad Re(\alpha_{ij})=0.
\eeq
For $a=k$, where $k=1,2,3$ we get
\beq
2Re\l{\chi}\r\gamma^k+2i\l{{\alpha^k}_0\gamma^0+{\alpha^k}_j\gamma^j}\r={\beta^k}_0\gamma^0+{\beta^k}_j\gamma^j,
\eeq
and hence
\beq
{\beta^j}_0=-2Im\l{{\alpha^j}_0}\r,\quad {\beta^i}_j=2Re\l{\chi}\r\delta^i_j-2Im\l{{\alpha^i}_j}\r.
\eeq
Recalling (\ref{znikanie}), we can finally conclude that 
\beq
{\beta^a}_b=\lambda\delta^a_b, \quad \mathbb{B}=\frac{1}{2}\lambda{\bf
  1}+i\mu_1{\bf 1}+i\mu_2\gamma^5,
\eeq
where $\lambda:=2Re\l{\chi}\r$, $\mu_1:=Im\l{\chi}\r$,
$\mu_2:=Im(\kappa)$ are real--valued one--forms.
Hence, the resulting new connection on $\mathcal{M}$ appears to be equal
\be\label{newc}
\stackrel{\lambda}{\omega}{^a}_b={\omega^a}_b+\lambda\delta^a_b.
\ee
Note the change of labeling -- instead of using $\mathbb{B}$, we are
now labeling this connection by
the real--valued one--form $\lambda$, on which it depends. Note also
that the one--forms $\mu_1$ and $\mu_2$ remain
undetermined, but they do not influence the connection
on the Riemann--Cartan space. If non--gravitational interactions were
considered, the components of these
one--forms could be hidden in the gauge fields corresponding to the
localization of the global symmetry of the change of phase $\psi\rightarrow e^{i\alpha}\psi$ and the
approximate symmetry under the chiral transformation $\psi\rightarrow
e^{i\alpha\gamma^5}\psi$. Since we do not want to deal here
with non--gravitational interactions, we will assume that $\mu_1$
and $\mu_2$ are equal to zero.\par
Let us investigate the properties of the new space--time connection
(\ref{newc}). It is no longer metric. The coefficients are expressed through
the coefficients of the original metric connection by
\be\label{conwsp}
\stackrel{\lambda}{\Gamma}{^{\rho}}_{\mu\nu}={\Gamma^{\rho}}_{\mu\nu}+\delta^{\rho}_{\mu}\lambda_{\nu}.
\ee
In a more mathematically oriented language, it could be defined by
\beq
\stackrel{\lambda}{\nabla}_Xf=X(f),\quad \stackrel{\lambda}{\nabla}_XY=\nabla_XY+\lambda(X)Y,
\eeq
where $f$ is a function and $X$, $Y$ are vector fields. Then the action on arbitrary tensor field is uniquely determined by
the requirements of commutativity with contraction and Leibniz
rule. The action on a $(r,s)$--tensorial type differential form is
given by
\beq
\stackrel{\lambda}{D}{T^{a_1\dots a_r}}_{b_1\dots b_s}=
D{T^{a_1\dots a_r}}_{b_1\dots b_s}+(r-s)\lambda\wedge {T^{a_1\dots a_r}}_{b_1\dots b_s}
\eeq 
and the action on a spinor field is
\beq
\stackrel{\lambda}{D}\psi=D\psi+\frac{1}{2}\lambda\psi.
\eeq
The torsion of this connection is
\be\label{Tors}
\stackrel{\lambda}{T}{^{cab}}=T^{cab}+\lambda^a\eta^{cb}-\lambda^b\eta^{ca}.
\ee
According to the leading idea, we do not want to introduce a new
field, but rather to express the one--form
$\lambda$ through the original connection $\omega$ in such a way that
the transformation rule for $\omega$ implies the appropriate
transformation rule for $\lambda$. According to (\ref{Btrans}), we
need to require that
$\mathbb{B}'=S(\Lambda)\mathbb{B}S^{-1}(\Lambda)$. But for
$\mathbb{B}=\frac{1}{2}\lambda{\bf 1}$ this simply means that
$\lambda$ is a scalar. The simplest possible construction which
fulfils the above mentioned requirements employs the torsion trace
one--form, $\mathbb{T}={T^{\nu}}_{\mu\nu}dx^{\mu}$. Let us denote by $\stackrel{+}{\nabla}$ and $\stackrel{-}{\nabla}$
the connections $\stackrel{\lambda}{\nabla}$ corresponding to the
choice $\lambda=\mathbb{T}$ and $\lambda=-\mathbb{T}$. The second of
them proved already useful in Einstein--Cartan theory. To see
this, note that (\ref{Tors}) enables one to rewrite the equation connecting torsion to the spin
distribution of matter (\ref{FEq}) in a particularly simple form
\beq
\stackrel{-}{T}{^{cab}}=k\tilde{S}^{abc}.
\eeq
What is more, one can prove the geometric identity
\beq
G^{ab}-G^{ba}=-\stackrel{-}{\nabla}_c\stackrel{-}{T}{^{cab}}.
\eeq
Together with (\ref{FEq}), these facts allow us to conclude, that 
\beq
\tilde{t}^{\,ab}-\tilde{t}^{\,ba}=\stackrel{-}{\nabla}_c\tilde{S}^{abc},
\eeq
if the field equations are satisfied. This formula corresponds to the
well--known relation between the canonical energy--momentum tensor and spin
density tensor in flat space,
$t^{\mu\nu}-t^{\nu\mu}=\partial_{\rho}S^{\mu\nu\rho}$.\par
The connection $\stackrel{+}{\nabla}$ has not yet been employed in
Einstein--Cartan theory. We wish to assign an even more significant
role to it. We postulate that this is the connection that ought to be
used in the procedure of minimal coupling. Then the flat space divergence of a
vector field will pass into 
\beq
\stackrel{+}{\nabla}_{\mu}V^{\mu}=\nabla_{\mu}V^{\mu}+T_{\mu}V^{\mu}={\stackrel{\circ}{\nabla}}_{\mu}V^{\mu},
\eeq
which becomes a differential when multiplied by the canonical
volume form $\epsilon$. Here we used (\ref{TV}) and
(\ref{conwsp}). Having the procedure corrected, we do not have to care
anymore about the choice of the flat space Lagrangian from the
class of equivalence defined by (\ref{Lch}). All of them will result
in the same theory after gravity is included. Unlike Saa's approach, our proposal is very conservative. As we will see
below, it leaves the predictions of the Einstein--Cartan theory almost
untouched.
\section{The consequences of the application of the new procedure.}\label{cons}
Having the procedure established, we can apply it to the Dirac
Lagrangian (\ref{LFR}). It is clear that the result will be the same
as the one that would be obtained by means of MCP, since
\beq
\stackrel{+}{D}\ov{\psi}\gamma^a\psi-\ov{\psi}\gamma^a\stackrel{+}{D}\psi=
D\ov{\psi}\gamma^a\psi-\ov{\psi}\gamma^aD\psi.
\eeq
Hence, all the predictions of the Einstein--Cartan gravity with fermions,
including the gravity--induced axial--axial point fermion interaction with a very
small coupling constant, remain valid. The advantage of the new
procedure is that we can use any other equivalent flat Lagrangian as a
starting point (in particular, we are now allowed to use the simplest one,
namely (\ref{LF0})). Hence, after the new procedure is adopted, the
vector--vector and axial--vector fermion interactions discussed in
\cite{Kazm1} do not appear anymore. The predictions of the theory
become unique and perhaps will enable experimental verification after
technological barriers are smoothed away in the future.\par
For a scalar field nothing will change,
as a covariant derivative of a scalar does not depend on whether we use
the connection $\nabla$ or $\stackrel{+}{\nabla}$.\par
Let us consider the vector field, described by the flat space Lagrangian density
\beq
\mathcal{L}_{vector}=-\frac{1}{4}F_{\mu\nu}F^{\mu\nu}+m^2A_{\mu}A^{\mu},
\quad F_{\mu\nu}=\partial_{\mu}A_{\nu}-\partial_{\nu}A_{\mu},
\eeq
where $m$ is a real parameter. For $m=0$, the Lagrangian is reduced to the
one of the electromagnetic field, whereas $m>0$ corresponds to Proca field.
One can define a one--form field $A=A_{\mu}dx^{\mu}$ and rewrite the derivative--dependent term as
\be\label{FA}
-\frac{1}{4}F_{\mu\nu}F^{\mu\nu}\epsilon=-\frac{1}{2}F\wedge\star F,
\quad F=dA=dA_{\mu}\wedge dx^{\mu}.
\ee
From the last equation it is clear that gravity can be included via
MCP in two different ways. One can either acknowledge 
$A$ as a scalar--valued one--form and assume that the exterior covariant
differential is just the usual differential, $DA=dA$, or assume that
$A_{\mu}$ is a one--form valued function and do the replacement
\beq
dA_{\mu}\wedge dx^{\mu}\rightarrow DA_a\wedge e^a, 
\quad DA_a=dA_a-{\omega^b}_aA_b.
\eeq
In the case of Maxwell field, the second procedure would lead to
the breaking of gauge invariance. Therefore, for Maxwell field, as well as
the non--Abelian gauge fields, the first possibility is usually
adopted. In the $m>0$ case both approaches are in principle possible,
however the latter one seems more appropriate, as it results in a
reasonable form of the spin density tensor\footnote{The first approach
  leads to vanishing of the dynamical spin density tensor
  $S^{abc}$ (see (\ref{FEq})).}. According to the ideas presented in this article,
nothing will change in the $m=0$ case (as well as in the case of
non--Abelian gauge fields). This is because the exterior covariant
differential of a scalar--valued one--form is just the usual
differential, independently of the choice of a connection. Hence, as
opposed to Saa's theory, the gauge fields do not couple to torsion in our
approach.
In the $m>0$ case, the usage of $\stackrel{+}{\nabla}$ instead of
$\nabla$ makes difference, as we end with the Lagrangian four--form
\beq
\ba
&\mathfrak{L}_{proca}=-\frac{1}{2}\stackrel{+}{F}\wedge\star \stackrel{+}{F}+m^2A_aA^a\epsilon,\\
&\stackrel{+}{F}=\tilde{F}-\mathbb{T}\wedge A,
\quad \tilde{F}=DA_a\wedge e^a=dA-A_aQ^a .
\ea
\eeq
Te dynamical spin density tensor is then 
\beq
\stackrel{+}{S}{^{abc}}=A^a\stackrel{+}{F}{^{cb}}-A^b\stackrel{+}{F}{^{ca}}+
A_d\l{\stackrel{+}{F}{^{ad}}\eta^{bc}-\stackrel{+}{F}{^{bd}}\eta^{ac}}\r,
\eeq
where $\stackrel{+}{F}{_{ab}}={\tilde F}_{ab}+A_aT_b-A_bT_a$,
${\tilde F}_{ab}=\tilde{e}_a(A_b)-\tilde{e}_b(A_a)-A_c{T^c}_{ab}$.
Had we used the usual MCP with the connection $\nabla$, we would have
obtained the spin density tensor
\beq
S^{abc}=A^a\tilde{F}^{cb}-A^b\tilde{F}^{ca}.
\eeq
Contrary to the Dirac field case, both the spin tensors depend on
torsion, which results in the second
equation of (\ref{FEq}) having a more complicated structure. It can
still be rewritten as a linear equation for the $24$ components of the
torsion tensor. However, the corresponding linear operator $L(A)$ may not be
invertible for some values of the field $A$, which leads to the
occurrence of the well--known torsion singularities \cite{HHKN} in the
theory. The determinant of this operator is 
$\det\l{L(A)}\r=2\l{2+kA_aA^a}\r^3$ in the case of standard MCP and 
$\det\l{L(A)}\r=16\l{1-2kA_aA^a}\r^3$ if the connection
$\stackrel{+}{\nabla}$ is used. These singularities are therefore not
removed by our procedure, but they are shifted towards other values of
$A$. It is surprising that the values of $A$ for which a singularity may appear are of opposite
causal character in these cases.
\section{Conclusions}\label{conc}
The modified coupling procedure introduced in Section \ref{MCA} provides a unique
method for coupling gravity to other field theories, which is
compatible with the Poincar{\'e} gauge description of the gravitational interaction. This method is a
slight modification of the procedure which could be considered as the
most natural choice for a gauge theory,
namely MCP. However, as opposed to MCP, the results obtained by this
method do not
depend on the choice of a flat space Lagrangian 
from the class of
equivalence connected with the possibility of the addition of a divergence. In particular,
this makes the predictions of EC gravity with fermions unique. They
appear to agree with those derived in earlier accounts for a
particular choice of fermionic Lagrangian. The classical theories of scalar field
and gauge fields in the presence of gravity do not change, if our
approach is adopted. For the massive vector field, the torsion
singularities appear to occur for different values of field than in
the standard EC treatment. This result is discussed
rather as a curiosity, as the set of fundamental matter fields which
are present in nature seems not to contain the Proca field.\par
There is finally an aesthetic argument which makes our procedure yet more
appealing. The fundamental motivations underlying the standard minimal
coupling procedure leave certain freedom, when applied to gravity. This
freedom is precisely employed by us to remove the ambiguity. In the
standard treatment of Poincar{\'e} gauge theory in which minimal
coupling procedure is used, the above mentioned freedom is ignored.
\section*{Acknowledgements}
I wish to thank Wojciech Kami{\'n}ski, Jerzy Lewandowski and Andrzej
Trautman for helpful comments and Urszula Pawlik for linguistic corrections. This work was partially supported by the Foundation for Polish
Science, Master grant.
\section{Appendix 1: Notation and conventions}\label{A1}
Throughout the paper $a,b,\dots$ are orthonormal tetrad indices and $\mu,\nu,\dots$ correspond to
a holonomic frame. For inertial frame of flat Minkowski space,
which is both holonomic and orthonormal, we use $\mu,\nu,\dots$.
The metric components in an orthonormal tetrad basis $\tilde{e}_a$ are
 $g\l{\tilde{e}_a,\tilde{e}_b}\r=(\eta_{ab})=diag(1,-1,-1,-1)$. Lorenz
 indices are shifted by $\eta_{ab}$. $\epsilon=e^0\wedge e^1\wedge
e^2\wedge e^3$ denotes the canonical
volume four--form  whose components in orthonormal tetrad basis obey $\epsilon_{0123}=-\epsilon^{0123}=1$.
The action of a covariant exterior differential $D$ on any $(r,s)$-tensorial
type differential $m$--form 
\beq
{T^{a_1\dots a_r}}_{b_1\dots b_s}=
\frac{1}{m!}{T^{a_1\dots a_r}}_{b_1\dots b_s\mu_1\dots \mu_m}d
x^{\mu_1}\w\dots\w d x^{\mu_m}
\eeq
is given by 
\beq
\ba
&D{T^{a_1\dots a_r}}_{b_1\dots b_s}:=d {T^{a_1\dots a_r}}_{b_1\dots b_s}\\
&+\sum_{i=1}^r{\omega^{a_i}}_c\w {T^{a_1\dots c\dots a_r}}_{b_1\dots
  b_s}
-\sum_{i=1}^s{\omega^c}_{b_i}\w {T^{a_1\dots a_r}}_{b_1\dots c\dots b_s} .
\ea
\eeq
The Hodge star action on external products of orthonormal cotetrad
one--forms is given by
\beq
\ba
&\star e_a=\frac{1}{3!}\epsilon_{abcd}e^b\w e^c\w e^d , \quad 
\star \l{e_a\w e_b}\r=\frac{1}{2!}\epsilon_{abcd}e^c\w e^d , \\ 
&\star \l{e_a\w e_b\w e_c}\r=\epsilon_{abcd}e^d ,
\ea
\eeq
which by linearity determines the action of $\star$ on any differential
form.
\section{Appendix 2: The principles of equivalence and general covariance in GR}\label{A2}
Note that {\it the principle of general covariance}, as formulated by
Weinberg on p.91 of \cite{WeinGr}, is not really
  of the same physical content as the {\it principle of equivalence},
  formulated on p.68, although Weinberg claims so. 
The latter allows for non--minimal couplings. Indeed, if
 we postulated that the motion of free falling material particle is
  determined by the equation 
\be\label{modgeod}
\frac{d^2x^{\mu}}{d\tau^2}+\alpha R{\Gamma^{\mu}}_{\rho\sigma}\frac{d
  x^{\rho}}{d\tau}\frac{d x^{\sigma}}{d\tau}+\beta R \frac{dx^{\mu}}{d\tau}=0,
\ee
where
$\tau(t)=\int_{t_0}^t\sqrt{\left|{g_{\mu\nu}\l{x(t)}\r\frac{dx^{\mu}}{dt}\frac{dx^{\nu}}{dt}}\right|}
dt$ is the proper time, $R$ the curvature scalar, $\Gamma$ represents
the Levi--Civita connection and $\alpha$,
$\beta$ are some real constants, which we shall call {\it non--minimal
  parameters}, then the principle of general covariance would not be
violated. What about the principle of equivalence? According to
Weinberg, it is formulated by the statement that {\it at every
  space--time point in an arbitrary gravitational field it is possible
  to chose a `locally inertial coordinate system' such that, within a
  sufficiently small region of the point in question, the laws of
  nature take the same form as in unaccelerated Cartesian coordinate
  systems in the absence of gravitation}. We will call this a weak
formulation of the principle of equivalence. Note that in a locally inertial
coordinate system the Christoffel symbols will vanish and therefore
the freedom of choice of the parameter $\alpha$ is not restricted by
such a principle. However, the parameter $\beta$ has
to be set to zero.

Now we wish to stress that in fact Weinberg used another, stronger
formulation of the equivalence principle to derive the predictions of
GR. The stronger principle states that the laws of physics governing
the behavior of matter and test particles in the
presence of gravity are {\it derivable} from those that are valid
in the absence of gravity by a well established rule. This
rule is that one rewrites the equations of special relativity,
originally given in inertial Cartesian coordinates, in arbitrary
coordinates. Then the connection coefficients, which assume nonzero
values in such coordinates even in flat space, have to be replaced by
the connection coefficients of the curved connection on the final
Riemannian manifold. In other words, in order to pass from flat to
curved space, one should merely replace all the derivatives by
the covariant ones. Such a prescription is identical with what we call the
minimal coupling procedure. Note that such a principle forces both the
parameters $\alpha$ and $\beta$ of (\ref{modgeod}) to vanish. It is
this strong formulation that enables one to derive the geodesic postulate,
as well as other concrete principles of GR, and finally makes the theory
experimentally verifiable.

\end{document}